\documentclass[11pt]{article}
\usepackage{vietnam,epsfig}

\begin{document}

\def\aj{Astron.\ J.}
\def\apj{ApJ}
\def\apjl{ApJ\ Lett.}
\def\apjs{ApJ\ Suppl.}
\def\aas{A\&A Supp.}
\def\aa{A\&A}
\def\aal{A\&A Lett.}
\def\mnras{MNRAS}
\def\mnrasl{MNRAS Lett.}
\def\nature{Nature}
\def\apss{Ap\&SS}
\def\pasp{{\it P.\ A.\ S.\ P.}}
\def\pasj{{\it PASJ}}
\def\pre{{\it Preprint}}
\def\aph{Astro-ph}
\def\adspr{{\it Adv. Space. Res.}}
\def\expas{{\it Experimental Astron.}}
\def\ssr{{\it Space Sci. Rev.}}
\def\ar{{\it Astronomy Reports}}

\def\ap{$\approx$ }
\def\mjysr{MJy/sr }
\def\mic{$\mu$m}


\title{Cosmology from Cosmic Microwave Background fluctuations with Planck}

\author{X. Dupac, on behalf of the Planck collaboration\footnote{Planck (www.rssd.esa.int/Planck) is an ESA project with instruments
funded by ESA member states (in particular the PI countries: France
and Italy), and with special contributions from Denmark and NASA.}}

\address{European Space Agency - ESAC, Villafranca del Castillo, Apdo. 50727, 28080 Madrid, Spain\\
Email: xdupac@sciops.esa.int}

\maketitle\abstracts{We present the main scientific goals and characteristics of the ESA Planck satellite mission, as well as the main features of the survey strategy and simulated performance in terms of measuring the temperature and polarization of the Cosmic Microwave Background fluctuations.
}

\section{Introduction}

In the last decade, there has been a large increase in the number of experiments dedicated to measurements of
the fluctuations of the Cosmic Microwave Background (CMB hereafter), e.g. Smoot et al. 1992, de Bernardis et al. 2000, Hanany et al. 2000, Kovac et al. 2002, Beno\^\i t et al. 2003, Pearson et al. 2003, Bennett et al. 2003.
These experiments seek to measure a
signal whose intensity is very low with respect to the background signal ($\sim 10^{-5}$), over a wide range of angular scales, with the most
ambitious experiments aiming at covering the whole sky with an angular resolution of the order of 10$'$ or better. Such
experiments are plagued by a host of systematic effects which arise in the instrument itself (e.g. 1/f detector noise, thermal effects), or in celestial signals which are picked up by the instrument (e.g. straylight due to
Galactic emission entering through far sidelobes).

\section{The Planck mission}

The Planck satellite (Tauber 2001) is a European Space Agency medium-size mission designed to map the whole sky at wavelengths ranging from 350 $\mu$m to 1 cm with unprecedented sensitivity (${\Delta T \over T} \approx 2 \times 10^{-6}$) and angular resolution (5$'$ - 30$'$). The instruments are tested and currently being integrated; the launch is foreseen in July 2008.

The Planck payload consists of a large offset 1.5-m telescope which collects radiation from the sky and
delivers it to a focal plane populated with 74 detectors.
These are of two kinds: tuned receivers based on low-noise amplifiers (HEMT) operating in the 20-80 GHz band (Low-Frequency Instrument, Mennella et al. 2003), and wideband bolometers operating in the 90-1000 GHz range (High-Frequency Instrument, Lamarre et al. 2003).
Both kinds of detector are very sensitive and stable, allowing them to operate in a ``total-power" mode.
This was not the case for the earlier CMB experiments which today define the state of the art of CMB anisotropy research (COBE/DMR and WMAP, http://lambda.gsfc.nasa.gov); instead they were designed to be differential to overcome inherent instabilities in the detectors.
The principle of Planck has however been demonstrated on smaller-scale balloon-based experiments, e.g. BOOMERANG (de Bernardis et al. 2000) and Archeops (Beno\^\i t et al. 2002).

The main observational objective of Planck is to image the temperature and polarization anisotropies of the CMB over the whole sky with uncertainties limited to natural causes such as cosmic variance and foreground fluctuations, rather than instrument noise, and with an angular resolution of 5' at high frequency channels.
Its main cosmological objective is to derive a very accurate angular power spectrum of the CMB fluctuations, both for the temperature and the E-mode polarization. There is also a possibility that it detects and analyzes B-mode CMB polarization signals.

\section{The Planck survey strategy}

In order to achieve a temperature accuracy per pixel of 10$^{-6}$ and provide the ability of measuring polarization (Stokes parameters I, Q, U) in the CMB channels with good cross-polar characteristics, extreme attention is put on controlling systematic effects:
\begin{itemize}
\item A wide frequency coverage is used, from 20 to 1000 GHz, which allows to control systematic effects and separate the astrophysical foreground components (Galactic dust, free-free, synchrotron, extragalactic sources, SZ galaxy clusters...) from the CMB
\item We use an orbit around the L2 Lagrange point of the Sun-Earth system, which allows to keep the perturbating bodies in terms of stray-light (Sun, Earth, Moon) always on the same side of the space-craft
\item We build the data redundancy on many time scales and angular scales, from one minute (the satellite spins at 1 rpm while keeping its spin axis fixed for about 45 min) to 6 months (after half a year the spin axis path gets back close to the initial position)
\item Details of the scanning strategy are optimized in order to ensure that the whole sky has been observed after about 8 months, and observed twice after 12 months, while minimizing the thermal disturbances on the space-craft and allowing relatively deep fields to exist around the Ecliptic poles
\end{itemize}

The baseline scanning strategy consists of roughly keeping the Planck spin axis in the anti-Sun direction, which means moving about one degree per day along the Ecliptic.
The optical axis follows the Planck rotation of 1 rpm with a bore-sight angle of 85 degrees, which allows to observe a large slice of the sky about 45 times while not moving the spin axis for about 45 min.
The exact motion of the spin axis along the year consists of following the anti-Sun plus a precession motion around the anti-Sun, with an amplitude of about 7$^o$ and a period of half a year.
The direction and phase of the precession are chosen to match telemetry-angle constraints (down-link antenna must point within 15$^o$ from the Earth) as well as to optimize the location of the deep fields on the sky with respect to dipole calibration and minimizing the amount of foregrounds in the deep fields.
More details about the Planck observation strategy can be found in Dupac \& Tauber (2005).

\section{Planck data processing}

The data processing is performed in two Data Processing Centers, one for each instrument, and consists in three main steps:
\begin{itemize}
\item Level 1: telemetry processing into time-ordered data
\item Level 2: calibration and map-making from time-ordered data, production of temperature and polarization frequency maps
\item Level 3: astrophysical component separation, production of component maps and CMB maps, production of angular power spectra
\end{itemize}

The Planck scientists will then analyze the maps and the power spectra to investigate a number of scientific areas, such as fundamental cosmology (determination of cosmological parameters, constraints on inflation and alternative theories, dark energy equation of state, etc), extragalactic science (submillimeter sources, galaxies, galaxy clusters with the Sunyaev-Zel'dovich effect...), Galactic science (structure, interstellar medium properties, etc) and Solar System science.

\section{Expected performances on cosmology}

The main objective of Planck concerns fundamental cosmology and the determination of the cosmological parameters through the temperature and E-mode polarization angula power spectra of the CMB fluctuations.
The two figures below show how Planck will improve the current knowledge in terms of maps and polarization angular power spectra.

\begin{figure}
\caption[]{Planck simulated CMB temperature maps of the whole sky in Mollweide projection (top), at COBE-DMR resolution (top left) and Planck resolution (top right); Planck simulated maps at high resolution (middle right) compared to WMAP (middle left); Planck CMB polarization maps (bottom). From the ``Scientific programme of Planck'' aka the ``Blue Book'', 2004.}
\label{mapss}
\end{figure}

\begin{figure}
\caption[]{Planck simulated E-mode polarization angular power spectrum (right) compared to WMAP and B2K ones (left). From the ``Blue Book'', 2004.}
\label{cl}
\end{figure}

The expected cosmological science to be performed from such results is very wide. We can cite a few domains of possible investigations:
\begin{itemize}
\item Determining the cosmological parameters within the current $\Lambda$CDM paradigm to high accuracy, without being indebted to other astrophysics or particle physics data
\item Probing causes of acceleration by dark energy (cosmological constant vs quintessence...)
\item Testing inflation models (non-gaussianity, etc)
\item Probing possible signatures of string theories through possible observable effects on the CMB polarization fluctuations
\end{itemize}

\section{Conclusion}

The Planck project has been designed to be the next milestone in CMB science. It should be able to provide unprecedented results on temperature and polarization anisotropies of the Cosmic Microwave Background, therefore allowing to determine with high precision the cosmological parameters, as well as putting new constraints on cosmology theories.


\end{document}